\documentclass[twocolumn,showpacs,preprintnumbers,amsmath,amssymb]{revtex4}
\usepackage{graphicx}
\usepackage{dcolumn}
\usepackage{bm}

\begin{document}
\title{Microscopic derivation of Open Quantum Walks}
\author{Ilya Sinayskiy}
 \email{sinayskiy@ukzn.ac.za}
  \affiliation{Quantum Research Group, School of Chemistry and Physics, University of KwaZulu-Natal, Durban, 4001, South Africa}
  \affiliation{National Institute for Theoretical Physics (NITheP), KwaZulu-Natal, South Africa}

\author{Francesco Petruccione}
 \email{petruccione@ukzn.ac.za}
  \affiliation{Quantum Research Group, School of Chemistry and Physics, University of KwaZulu-Natal, Durban, 4001, South Africa}
  \affiliation{National Institute for Theoretical Physics (NITheP), KwaZulu-Natal, South Africa}

\date{\today}

\begin{abstract}
Open Quantum Walks (OQWs) are exclusively driven by dissipation and are formulated as completely positive trace preserving (CPTP) maps on underlying graphs. The microscopic derivation of discrete and continuous in time OQWs is presented. It is assumed that connected nodes are weakly interacting via a common bath. The resulting reduced master equation of the quantum walker on the lattice is in the generalised master equation form. The time discretisation of the generalised master equation leads to the OQWs formalism. The explicit form of the transition operators establishes a connection between dynamical properties of the OQWs and thermodynamical characteristics of the environment. The derivation is demonstrated for the examples of the OQW on a circle of nodes and on a finite chain of nodes. For both examples a transition between diffusive and ballistic quantum trajectories is observed and found to be related to the temperature of the bath.
\end{abstract}

\maketitle

\section{Introduction}
The mathematical concept of classical random walks (CRW) finds wide applications in various areas of fundamental and applied science \cite{pt,rwp,rwcs,rwe,rwb}. In the case of the classical random walk the trajectory of the ``walker" is a sequence of random steps fully determined by the stochastic matrix corresponding to the graph of the CRW. Unitary quantum walks have been introduced more than two decades ago, as the quantum analogue of the CRW  \cite{aharonov, kempe, qwrev}. Unitary quantum walks are broadly used in quantum computing science, in the formulation of the quantum algorithms and in complexity theory \cite{qwrev}. The dynamical behaviour of the quantum walker and the classical random walker are very different. The probability distribution of the unitary quantum walker is the result of the interference between different positions of the walker and it is determined not only by the underlying graph, but also by the inner state of the quantum walker, e.g., spin or polarisation \cite{qwrev}.

In the description of the dynamics of any realistic quantum system one needs to take into account the effect of dissipation and decoherence \cite{toqs}. In the recently introduced open quantum walks (OQWs) these effects are naturally included into the description of the dynamics of the quantum ``walker" \cite{pla,JSP,jpc}. Essentially, the dynamics of OQWs is driven by the dissipative interaction with environments. Mathematically OQWs are formulated as completely-positive trace preserving maps (CPTP maps) on an appropriate Hilbert space \cite{toqs,kraus}. OQWs and unitary quantum walks can be related via a ``physical realisation" procedure introduced in Ref. \cite{JSP}. In a special scaling limit OQWs become Open Quantum Brownian Motion \cite{bm1,bm2}. This is a new type of quantum Brownian motion where the position of the quantum Brownian particle is determined not only by the interaction with an environment but also by the state of the inner degree of freedom of the quantum Brownian particle. 

The diverse dynamical behaviour of OQWs has been extensively studied \cite{pla,JSP,jpc,anal,doqw,OQW12,OQW13,OQW11,Clem}. The asymptotic analysis of OQWs leads to a central limit theorem \cite{attal3,konno,clt01}. For large times the position probability distribution of OQWs converges to Gaussian distributions \cite{attal3,konno}. For a special case of OQWs on $\mathbb{Z}$ it was shown that the probability distribution of the position of the quantum walker is given by the binomial distribution and the number of different Gaussian distribution for large times is bound by the number of inner degrees of  freedom of the OQWs \cite{pscp}.

OQWs can perform dissipative quantum computing (DQC) \cite{verstraete} and quantum state engineering \cite{qip}. It has been shown that the OQW implementation of DQC outperforms the conventional model of DQC. In particular, it has been found that OQWs can be designed to converge faster to the desired steady state and to increase the probability of detection of the outcome of the quantum computation  \cite{qip}.

A quantum optical implementation of simple OQWs has been suggested using a dissipative out of resonance cavity QED setup  \cite{ijqi}. The Fock states of the cavity mode correspond to the nodes of the walk and the state of the two-level system corresponds to an inner degree of freedom of the walker.

Recently, Bauer \emph{et al.} found that the OQW quantum trajectories  can switch between diffusive and ballistic behaviour \cite{bm1}. However, such switching was established for abstract CPTP maps without identifying the physics of the underlying system.

OWQs have been introduced formally and the question of microscopic models leading to OQWs needs to be addressed. Due to the dissipative nature of the OQWs it is natural to expect them to be obtained from an appropriate system-environment model as the reduced dynamics of a quantum walker on a graph. Indeed, for a simple case of an OQW on a two-node graph a microscopic derivation has been described \cite{osid}.

The aim of the present paper is to derive OQWs from a microscopic Hamiltonian on an arbitrary graph. The microscopic derivation includes the identification of an appropriate system environment model (Hamiltonian of the system, bath and system-bath interaction), the Born-Markov approximation and tracing out the environmental degrees of freedom \cite{toqs,gard,carm}. The resulting quantum master equation will be shown to have the form of a generalised master equation \cite{breuer} and a discrete time version corresponds to the CPTP maps defining OQWs. The formalism is demonstrated for two examples of OQWs, namely an OQW on a circle of nodes and an OQW on a finite chain of nodes. The microscopic derivation allows to connect the thermodynamical parameters of the environment and the dynamics of OQWs. It is shown that the temperature of the environment plays the role of the switch between diffusive and ballistic quantum trajectories corresponding to the OQWs.

The paper has the following structure. In Section II we briefly review the formalism of OQWs. In Section III we formulate the microscopic model and perform the microscopic derivation of the OQWs. In Section IV we demonstrate the microscopic derivation for two examples. In Section V we conclude.

\section{Formalism of Open Quantum Walks}

\begin{figure}
\includegraphics[width= \linewidth]{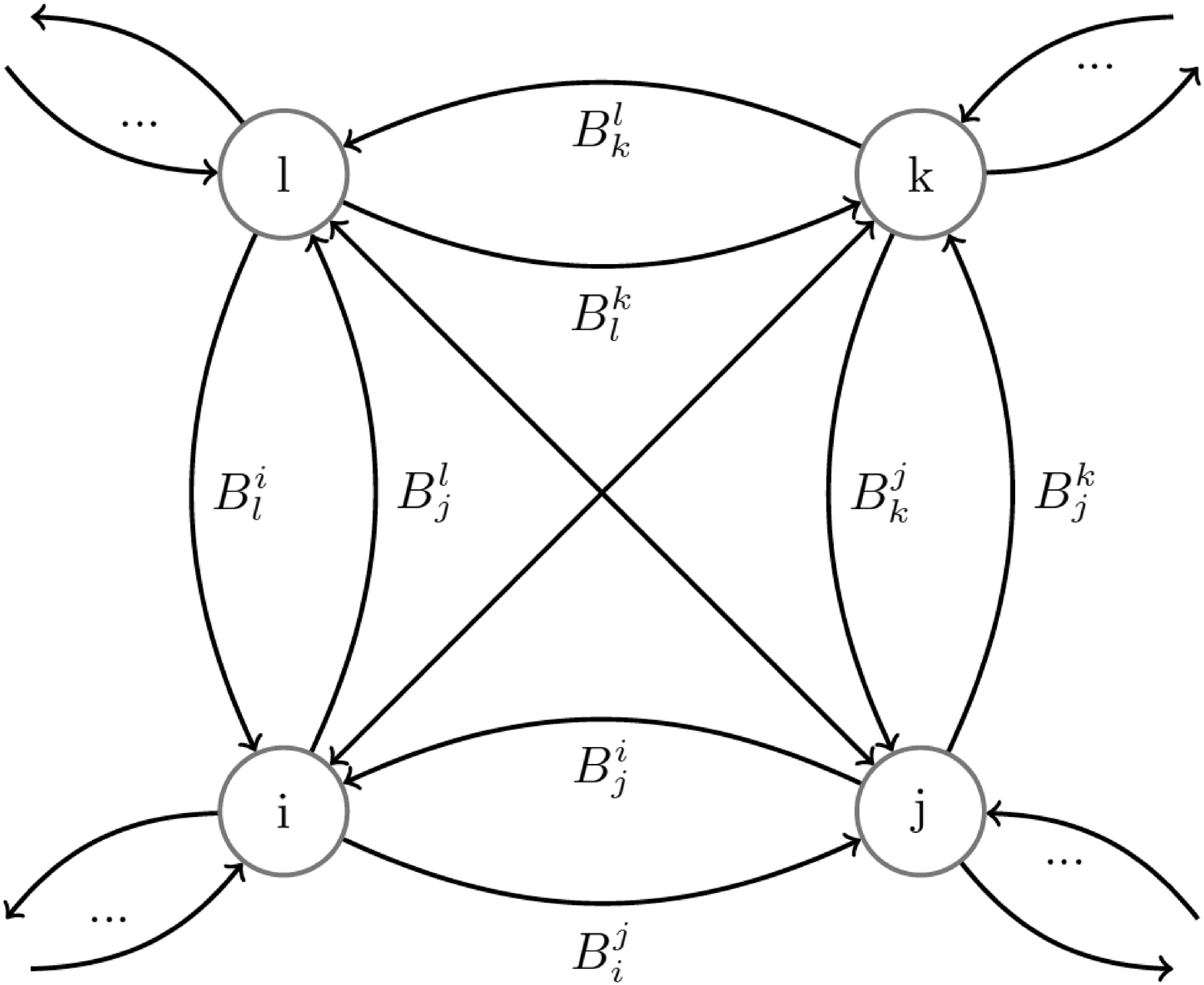}
\caption{Schematic illustration of the formalism of Open Quantum Walks. The walk is realised on a graph with a set of vertices denoted by $i,j,k,l$.  The operators $B_i^j$ describe transformations in the internal degree of freedom of the ``walker" during the transition from node $(i)$ to node $(j)$.}
\end{figure}

The open quantum walk is a quantum walk defined on a set of nodes, where the transitions between the nodes are exclusively driven by dissipation. Mathematically, open quantum walks are formulated in the language of completely positive trace preserving (CPTP) maps. A CPTP map is defined on the graph $\mathcal{G}=\left(\mathcal{V},E\right)$, where $\mathcal{V}$ is the set all nodes and $E=\{(i,j)|i,j\in\mathcal{V}\}$ is the set of all ordered edges denoting possible transitions between the nodes. The number of nodes can be finite $(P<\infty)$ or infinitely countable $(P=\infty)$, where $P$ denotes cardinality of the set of nodes $\mathcal{V}$, i.e.  $P=\mathrm{card}(\mathcal{V})$. The corresponding Hilbert space of the walk is defined as separable Hilbert space $\mathcal{K}=\mathbb{C}^{P}$ for the finite-dimentional case $(P<\infty)$ and for the infinite-dimentional case $(P=\infty)$ the Hilbet space is the space of square integrable functions $\mathcal{K}=l^2(\mathbb{C})$ with orthonormal basis indexed by $|i\rangle$ where $i\in \mathcal{V}$. The internal degrees of freedom of the quantum walker, e.g. spin, polarisation or $n$-energy levels, are described by a separable Hilbert space $\cal{H}_N$ attached to each node. Any state of the walker will be described on the direct product of the Hilbert spaces $\cal{H}_N\otimes \cal{K}$. 
 
In order to describe the dynamics of the internal degree of freedom of the walker for each edge $(i,j)$ we introduce bounded operators $B^i_j$ acting on $\cal{H}_N$. These operators describe the transformation of the internal degree of freedom of the quantum walker due to the ``jump" from node $j$ to node $i$ (see Fig. 1). On each node $j$ we define a CPTP map $\begin{mathcal}M\end{mathcal}_j$ in the Kraus representation on the space of operators on $\mathcal{H}_N$,
\begin{equation}
\begin{mathcal}M\end{mathcal}_j(\tau)=\sum_i B^i_j \tau {B^i_j}^\dag.
\end{equation}
Complete positivity and trace preservation of the above map is guaranteed by the following normalisation condition for each node $j$ \cite{kraus},
\begin{equation}\label{eq1}
\sum_i {B^i_j}^\dag B^i_j= I.
\end{equation}
This condition is the generalisation of the classical Markov chain condition.

From the physical point of view, the operators $B^i_j$  affect only the internal degrees of freedom of the walker and do not perform transitions from node $j$ to node $i$. We can extend the 
 action of the operators $B^i_j$ on the whole lattice with the help of the following dilation,
\begin{equation}
M^i_j=B^i_j\otimes | i\rangle\langle j|\,.
\end{equation}
If the condition expressed in Eq. (\ref{eq1}) is satisfied, then $\sum_{i,j} {M^i_j}^\dag M^i_j=1$ \cite{JSP}. This normalisation condition allows to define a CPTP map for density matrices on $\mathcal{H}_N\otimes\mathcal{K}$, i.e.,
\begin{equation}\label{OQW}
\begin{mathcal}M\end{mathcal}(\rho)=\sum_i\sum_j M^i_j\,\rho\, {M^i_j}^\dag.
\end{equation}
The CPTP-map $\begin{mathcal}M\end{mathcal}$ defines the discrete time \textit{Open Quantum Walk} \cite{pla, JSP}. It has been shown that for an arbitrary initial state the density matrix $\sum_{i,j} \rho_{i,j}\otimes| i\rangle\langle j|$ will take a diagonal form in position Hilbert space $\cal{K}$ after just one step of the OQW \cite{pla, JSP}. For this reason it is sufficient to assume that the initial state of the system is in the diagonal form in the ``position" space $\rho=\sum_i \rho_i\otimes | i\rangle\langle i|$ with $\sum_i { \mathrm Tr}(\rho_i)=1$.

It is straightforward to give an explicit iteration formula for the OQW from step $n$ to step $n+1$ 
\begin{equation}\label{rho}
\rho^{[n+1]}=\begin{mathcal}M\end{mathcal}(\rho^{[n]})=\sum_i \rho_i^{[n+1]}\otimes | i\rangle\langle i|,
\end{equation}
where
\begin{equation}
\rho_i^{[n+1]}=\sum_j B^i_j \rho_j^{[n]}{B^i_j}^\dag.
\end{equation}
This iteration formula gives a clear physical meaning to the CPTP-mapping that we introduced: the state of the system on node $i$ is determined by the conditional shift from all connected nodes $j$  and the internal state of the walker on the node $j$. These conditional shifts are defined by the explicit form of the operators $B_j^i$.  Also, it is straightforward to see that $\mathrm{Tr}[\rho^{[n+1]}]=\sum_i\mathrm{Tr}[\rho_{i}^{[n+1]}]=1$.

As an example let us consider a homogeneous OQW on the circle or on the line (see Fig. 2) with jump operators $B_j^i$ defined as
\begin{equation}
B_{i}^{i+1}\equiv B, \quad B_i^i\equiv A, \quad B_{i}^{i-1}\equiv C
\end{equation}
and the normalisation condition given by 
$A^\dag A+B^\dag B+C^\dag C=1$.
The state of the walker $\rho^{[n]}$ after $n$ steps reads,
\begin{equation}
\rho^{[n]}=\sum_i \rho_i^{[n]}\otimes|i\rangle\langle i|,
\end{equation}
where the form of the $\rho_i^{[n]}$ $(i\in \mathbb{Z})$ is easily found by iteration,
\begin{equation}
\label{eq:iteration}
\rho_i^{[n+1]}=A\rho_{i}^{[n]}A^\dag+B\rho_{i-1}^{[n]}B^\dag+C\rho_{i+1}^{[n]}C^\dag.
\end{equation}
Several examples of OQWs on $\mathbb{Z}$ can be found in  \cite{pla,JSP,jpc,pscp}.

\section{Microscopic derivation of OQWs}

As it was stated before, OQWs are formulated as quantum walks on a set of nodes, where the transition between the nodes are driven by dissipation. This implies that it should be possible to derive OQWs using methods of the theory of open quantum systems \cite{toqs}. From a microscopic point of view the Hamiltonian of the total system is given by
\begin{equation}
H=H_S+H_B+H_{SB},
\end{equation}
where in the usual notation $H_S$, $H_B$ and $H_{SB}$ stand for Hamiltonian of the system, bath and system-bath interaction, respectively.
As the result of the microscopic derivation we would like to obtain OQWs after tracing out the bath degrees of freedom. In the context of this work, the open system is not only the ``quantum walker", but the quantum walker with the underlying lattice. Using the condition that the transitions between the nodes are driven by dissipation, it is easy to conclude that the system Hamiltonian by itself should describe only the local dynamics of the walker on each node which implies the following form of the system Hamiltonian $H_S$ for the $M$-node network,
\begin{equation}
H_S=\sum_{i=1}^M\Omega_i\otimes|i\rangle\langle i|.
\end{equation}
The position of the walker is described by the set of orthogonal vectors $\{|i\rangle\}_{i=1}^M$ which form the basis in the position Hilbert space $\mathcal{C}^M$. The state of the inner degree of freedom of the walker is described by the operators $\Omega_i\in\mathcal{B}(\mathcal{H}_N)$, where $\mathcal{H}_N$ is an $N$-dimensional Hilbert space describing the inner degrees of freedom of the walker. 

OQWs are designed such that the transitions between different sites are uncorrelated. This means that for each pair of nodes $i$ and $j$ between which transitions are possible one needs to have at least one local environment which will drive the walker between the nodes. The direct consequence of this is the following form of the bath Hamiltonian $H_B$:
\begin{equation}
H_B=\sum_{i\neq j=1}^M\sum_n \omega_{i,j,n}a_{i,j,n}^\dag a_{i,j,n},
\end{equation}
where $a_{i,j,n}^\dag$ and $a_{i,j,n}$ are bosonic creation and annihilation operators of the modes (photonic/phononic) of the bath, with standard commutation relations.

The system-bath Hamiltonian $H_{SB}$ describes environment assisted transitions of the quantum walker between the nodes. This implies that for any two nodes $i$ and $j$ this Hamiltonian has the following structure $H_{SB}^{i\leftrightarrow j}=A_{i,j}\otimes X_{i,j} \otimes B_{i,j}$. As in a typical open quantum system we assume a linear coupling between each component of the total system, but the crucial difference is that in the present work we have two different degrees of freedom of the system, namely the position and the internal degree of freedom, coupled simultaneously to the bath. The operator $A_{i,j} \in\mathcal{B}(\mathcal{H}_N)$ is an operator acting on the internal degree of freedom of  the  walker and plays the role of a ``quantum coin" similar to the Hadamard matrix for the unitary quantum walks \cite{kempe,qwrev}. The operator $A_{ij}$ is conditioning the probability of transition between the nodes to the internal state of the walker. If the operator $A_{ij}$ is trivial, i.e. $A_{i,j}\equiv 1$, then the resulting open walk will be classical. The operator $X_{i,j}\in\mathcal{B}(\mathcal{C}^M)$ describes the transition between the nodes $i$ and $j$, the simplest choice is the following, $X_{i,j}=|j\rangle\langle i|+|i\rangle\langle j|$. The coupling between the quantum walker and the corresponding environment is described by the operator $B_{i,j}$, the simplest and typical choice is linear coupling of the system to the bath, i.e. $B_{i,j}=\sum_n g_{i,j,n}a_{i,j,n} +g_{i,j,n}^*a_{i,j,n}^\dag$. It is clear, that only system-bath Hamiltonian containing the tensor product of these three operators $(A_{i,j}, X_{i,j}, B_{i,j})$ is minimally required to obtain OQWs. As it was stated above if $A_{ij}$ is trivial, the walk will be classical, if $X_{i,j}$ is trivial there will be no walk and finally if $B_{ij}$ is trivial the quantum walk would not be environment driven. The system bath Hamiltonian $H_{SB}$ describing environment assisted transitions has the following form,
\begin{equation}
H_{SB}=\sum_{i,j}\sum_n A_{i,j}\otimes X_{i,j}\otimes\left(g_{i,j,n}a_{i,j,n} +g_{i,j,n}^*a_{i,j,n}^\dag\right).
\end{equation}
Having specified the Hamiltonian of the total system, we can proceed and derive the reduced master equation describing the dynamics of the system which consists of the quantum walker and the lattice. Here we assume that the system is weakly interacting with the reservoirs, so that the Born-Markov approximation is valid \cite{toqs}. Under these assumptions the reduced dynamics of the system in the interaction picture is given by the following equation,
\begin{eqnarray}
\label{eq:BME}
& &\frac{d}{dt}\rho_s(t)=\\\nonumber
&-&\int_0^\infty d\tau \mathrm{Tr}_B\left[H_{SB}(t),\left[H_{SB}(t-\tau),\rho_s(t)\otimes\rho_B\right]\right],
\end{eqnarray}
where $\rho_s(t)$ is the reduced density matrix of the system (walker on the network) and $\rho_B$ is the state of the reservoir. In general, Eq. (\ref{eq:BME}) does not guarantee that the resulting master equation will be in the Gorini-Kossakowski-Sudarshan-Lindblad form (GKSL-form) \cite{toqs,GKS,L,davies}. To obtain, the master equation in the form of the generator of the dynamical semigroup one needs to perform an additional rotating wave approximation \cite{toqs}. This rotating wave approximation can be straightforwardly performed if we decompose the system-bath Hamiltonian in the basis of eigenoperators of the system Hamiltonian $H_S$. To this end on each node $|i\rangle$ we introduce the set of orthonormal projection operators $\{\Pi_i(\lambda^{(i)})\}$  onto the eigenvalues $\lambda^{(i)}$ of each Hamiltonian $\Omega_i$ such that,
\begin{equation}
\Omega_i=\sum_{\lambda^{(i)}}\lambda^{(i)}\Pi_i(\lambda^{(i)}).
\end{equation}
It is easy to see that in the interaction picture the system-bath Hamiltonian $H_{SB}$ takes now the following form,
\begin{widetext}
\begin{equation}
H_{SB}(t)=\sum_{i,j}\sum_{\lambda^{(i)},\lambda^{(j)}}e^{it(\lambda^{(i)}-\lambda^{(j)})}\Pi_i(\lambda^{(i)})A_{i,j}\Pi_j(\lambda^{(j)})\otimes |i\rangle\langle j|\otimes B_{i,j}(t)+\mathrm{h.c.},
\label{EQ:HSB}
\end{equation}
\end{widetext}
where the operator $B_{i,j}(t)$ is given by,
\begin{eqnarray}\nonumber
B_{i,j}(t)&=&e^{itH_B}\left(\sum_n g_{i,j,n}a_{i,j,n} +g_{i,j,n}^*a_{i,j,n}^\dag\right)e^{-itH_B}\\
&=&\sum_n g_{i,j,n}a_{i,j,n}e^{-it\omega_{i,j,n}} +\mathrm{h.c.}.
\end{eqnarray}
Eq. (\ref{EQ:HSB}) can be rewritten as,
\begin{eqnarray}
H_{SB}(t)&=&\sum_{i,j}\sum_{\omega}e^{it\omega}A_{i,j}^\dag(\omega)\otimes |i\rangle\langle j|\otimes B_{i,j}(t)+\mathrm{h.c.}\\\nonumber
&+&\sum_{i,j}\sum_{\omega'}e^{-it\omega'}A_{i,j}(\omega')\otimes |i\rangle\langle j|\otimes B_{i,j}(t)+\mathrm{h.c.},
\end{eqnarray}
where the operators $A_{i,j}^\dag(\omega)$ and $A_{i,j}(\omega')$ are defined as,
\begin{eqnarray}\nonumber
A_{i,j}(\omega) &=& \sum_{\lambda^{(i)}-\lambda^{(j)}=\omega< 0}\Pi_i(\lambda^{(i)})A_{i,j}\Pi_j(\lambda^{(j)}),\\
A_{i,j}^\dag(\omega')&=& A_{i,j}(-\omega').
\end{eqnarray}
It is straightforward to see that the summation over all frequencies $\omega$ and $\omega'$ gives the original operators $A_{i,j}$,
\begin{equation}
\sum_{\omega}A_{i,j}(\omega)+\sum_{\omega'}A_{i,j}^\dag(\omega')=A_{i,j}.
\end{equation} 
Having defined the system-bath Hamiltonian $H_{SB}$, we can put the explicit expression for the Hamiltonian Eq. (\ref{EQ:HSB}) into the generic equation for the reduced density matrix Eq. (\ref{eq:BME}) and obtain the master equation for the system. In Eq. (\ref{eq:BME}) we will need to trace out the bath degrees of freedom. Here, we assume that the environment is in a thermal equilibrium state, i.e., the state of the reservoir is given by the canonical distribution $\rho_B=\exp{(-\beta H_B)}/\mathrm{Tr}[\exp{(-\beta H_B)}]$, where $\beta$ is the inverse temperature of the bath $\beta=(k_BT)^{-1}$.
Using the explicit form of the system-bath interaction Hamiltonian in the interaction picture $H_{SB}(t)$ Eq. (\ref{EQ:HSB}) and utilising the rotating wave approximation for the transition frequencies $\omega$ and $\omega'$ \cite{toqs,davies} it follows,
\begin{eqnarray}
\label{EQ:DMIJ}\nonumber
\frac{d}{dt}\rho_s(t)&=&\sum_{i,j}\sum_{\omega}\gamma_{i,j}(-\omega)\mathcal{L}\left(A_{i,j}(\omega)\otimes|j\rangle\langle i|\right)\rho_s(t)\\\nonumber
& &+\gamma_{i,j}(\omega)\mathcal{L}\left(A_{i,j}^\dag(\omega)\otimes|i\rangle\langle j|\right)\rho_s(t)\\\nonumber
& &+\sum_{i,j}\sum_{\omega'}\gamma_{i,j}(-\omega')\mathcal{L}\left(A_{i,j}(\omega')\otimes|i\rangle\langle j|\right)\rho_s(t)\\
& &+\gamma_{i,j}(\omega')\mathcal{L}\left(A_{i,j}^{\dag}(\omega')\otimes|j\rangle\langle i|\right)\rho_s(t),
\end{eqnarray}
where $\mathcal{L}\left(A\right)\rho$ denotes the dissipative superoperator in diagonal GKSL-form \cite{toqs, GKS, L},
\begin{equation}
\label{eq:GKSL}
\mathcal{L}\left(A\right)\rho=A\rho A^\dag-\frac{1}{2}\left\{A^\dag A,\rho\right\}
\end{equation}
and $\gamma_{i,j}(\omega)$ is the real part of the Fourier transform of the reservoir correlation functions  $\langle B_{i,j}^\dag(\tau)B_{i,j}(0)\rangle$,
\begin{equation}
\gamma_{i,j}(\pm\omega)=\frac{\gamma_{i,j}^{\mathrm{se}}}{2}\left(\coth\left(\frac{\beta\omega}{2}\right)\mp1\right),
\end{equation}
where $\gamma_{i,j}^{\mathrm{se}}$ is the coefficient of the spontaneous emission in the corresponding local reservoir. In Eq. (\ref{EQ:DMIJ}) the Lamb-type shift terms are neglected. These terms describe shifts in energy levels of the system due to the interaction with the heat bath and typically do not influence the dissipative dynamics of the system. The value of the Lamb-type shift is much smaller than other characteristic parameters in the system Hamiltonian and traditionally these terms are dropped. For the cases when the dimension of the reduced system is large enough these Lamb-type shifts might affect the dissipative dynamic of the reduced system \cite{NJP, basharov}. However, for simplicity in the present manuscript we assume that one can neglect these contributions.

It is interesting to note that the quantum master equation Eq. (\ref{EQ:DMIJ}) has the form of the generalised master equation, introduced by Breuer \cite{breuer}. For the generalised master equation it is assumed that the total density matrix of the reduced system can be written as $\rho=\sum_i \rho_i\otimes|i\rangle\langle i|$, where each operator $\rho_i$ satisfies the following differential equation,
\begin{equation}
\frac{d}{dt}\rho_i=\mathcal{K}_i\left(\rho_1,..,\rho_n\right),
\end{equation}
where  the general form of the generators $\mathcal{K}_i$ is given by
\begin{eqnarray}
\mathcal{K}_i\left(\rho_1,...,\rho_n\right)&=&-i[H_i,\rho_i]\\\nonumber
&+&\sum_{j,\lambda}\left(R_\lambda^{ji}\rho_jR_\lambda^{ji\dag}-\frac{1}{2}\{R_\lambda^{ji\dag}R_\lambda^{ji},\rho_i\}\right),
\end{eqnarray} 
with Hermitian operators $H_i$ and non-Hermitian operators $R_\lambda^{ji}$.

It is straightforward to see, that writing the density matrix of the reduced system from Eq. (\ref{EQ:DMIJ}) as $\rho_s(t)=\sum_{i=1}^M \rho_i(t)\otimes|i\rangle\langle i|$, where $|i\rangle\langle i|$ is a projection on the node $i$, then the quantum master equation (\ref{EQ:DMIJ}) reduces to the system of differential equations for $\rho_i(t)$:

\begin{equation}
\label{EQ:DMIJ1}
\frac{d}{dt}\rho_i(t)= \mathcal{K}_i\left(\rho_1,...,\rho_M\right),
\end{equation}
where $\mathcal{K}_i\left(\rho_1,...,\rho_M\right)$ is now explicitly given by
\begin{widetext}
\begin{eqnarray}\label{ctoqw}
\mathcal{K}_i\left(\rho_1,...,\rho_M\right)&=&\sum_{j,\omega}\gamma_{j,i}(-\omega)A_{j,i}\left(\omega\right)\rho_jA_{j,i}^\dag\left(\omega\right)-\frac{\gamma_{i,j}(-\omega)}{2}\{A_{i,j}^\dag \left(\omega\right)A_{i,j}\left(\omega\right),\rho_i\}\\\nonumber
&+&\sum_{j,\omega}\gamma_{i,j}(\omega)A_{i,j}^\dag\left(\omega\right)\rho_jA_{i,j}\left(\omega\right)-\frac{\gamma_{j,i}(\omega)}{2}\{A_{j,i}\left(\omega\right)A_{j,i}^\dag\left(\omega\right),\rho_i\}\\\nonumber
&+&\sum_{j,\omega'}\gamma_{i,j}(-\omega')A_{i,j}\left(\omega'\right)\rho_jA_{i,j}^\dag\left(\omega'\right)-\frac{\gamma_{j,i}(-\omega')}{2}\{A_{j,i}^\dag\left(\omega'\right) A_{j,i}\left(\omega'\right),\rho_i\}\\\nonumber
&+&\sum_{j,\omega'}\gamma_{j,i}(\omega')A_{j,i}^\dag\left(\omega'\right)\rho_jA_{j,i}\left(\omega'\right)-\frac{\gamma_{i,j}(\omega')}{2}\{A_{i,j}\left(\omega'\right)A_{i,j}\left(\omega'\right)^\dag,\rho_i\}.\\\nonumber
\end{eqnarray}
\end{widetext}
The system of differential equations (\ref{EQ:DMIJ1}) and (\ref{ctoqw}) defines the \textit{continuous time open quantum walk}. Continuous time OQWs have been introduced recently, as  continuos in time limit of the discrete time OQWs Eq. (\ref{rho})  \cite{Clem}.

In order to obtain a discrete time OQW in the form (\ref{OQW}), one needs to introduce discretised time steps. There are at least two ways of achieving this. The first way is to discretise the solution of the system of equations  (\ref{EQ:DMIJ1}), however, it cannot be done in a generic setting. The second way, which is the one way to do a discretisation in general, is to discretise the system of the differential equations (\ref{EQ:DMIJ1}). To do this one needs to replace the time derivative by the finite difference with a small time step $\Delta$,
\begin{equation}
\frac{d}{dt}\rho_i(t)\rightarrow \frac{\rho_i(t+\Delta)-\rho_i(t)}{\Delta}.
\end{equation}
The above substitution leads to the following transition operators,
\begin{widetext}
\begin{eqnarray}
\label{genJump}
B_j^{i(1)}(\omega)=\sqrt{\Delta\gamma_{j,i}(-\omega)}A_{j,i}(\omega),\quad B_j^{i(2)}(\omega)=\sqrt{\Delta\gamma_{i,j}(\omega)}A_{i,j}^\dag(\omega),\\\nonumber
B_j^{i(1)}(\omega')=\sqrt{\Delta\gamma_{i,j}(-\omega')}A_{i,j}(\omega'),\quad B_j^{i(2)}(\omega')=\sqrt{\Delta\gamma_{j,i}(\omega')}A_{j,i}^\dag(\omega'),\\\nonumber
B_i^i=I_N-\frac{\Delta}{2}\sum_{j,\omega}\left(\gamma_{i,j}(-\omega)A_{i,j}^\dag(\omega)A_{i,j}(\omega)+\gamma_{j,i}(\omega)A_{j,i}(\omega)A_{j,i}^\dag(\omega)\right)\\\nonumber
-\frac{\Delta}{2}\sum_{j,\omega'}\left(\gamma_{j,i}(-\omega') A_{j,i}^\dag(\omega')A_{j,i}(\omega')+\gamma_{i,j}(\omega') A_{i,j}(\omega')A_{i,j}(\omega')\right),
\end{eqnarray}
\end{widetext} 
where $I_N$ is an $N$-dimensional identity operator on the Hilbert space $\mathcal{H}_N$. One can see that the set of transition operators introduced above satisfies standard normalisation conditions up to $\mathcal{O}(\Delta^2)$,
\begin{eqnarray}
B_{j}^{j\dag}B_{j}^{j}&+&\sum_{k=1}^2\sum_{j,i,\omega}B_j^{i(k)\dag}(\omega)B_j^{i(k)}(\omega)\\\nonumber
&+&\sum_{k=1}^2\sum_{j,i,\omega}B_j^{i(k)\dag}(\omega')B_j^{i(k)}(\omega')=I_N.
\end{eqnarray}
Hence, the iteration formula for the discrete time OQW reads,
\begin{eqnarray}
\label{OQWgen}
\rho_i^{[n+1]}=B_i^i\rho_i^{[n]}B_i^{i\dag}+\sum_{k=1}^2\sum_{j,\omega}B_j^{i(k)}(\omega)\rho_j^{[n]}B_j^{i(k)\dag}(\omega)\\\nonumber
+\sum_{k=1}^2\sum_{j,\omega'}B_j^{i(k)}(\omega')\rho_j^{[n]}B_j^{i(k)\dag}(\omega').
\end{eqnarray} 

The explicit expression for the transition operators Eq. (\ref{genJump}) establishes a connection between the dynamical properties of the OQWs and the thermodynamical parameters of the environment.

\section{Examples of Open quantum walks}

\subsection{First example}
As an example of the above microscopic derivation of OQWs, we consider a two-level system (two-level atom or electron with spin) as a quantum walker on a circle with $M$ nodes (Fig. 2). In this case the system Hamiltonian $H_S$ of the two level system (``quantum walker") on the circle reads,
\begin{equation}
H_S=\sum_{i=1}^M\frac{\omega_0}{2}\sigma_z\otimes|i\rangle\langle i|+\lambda(\vec{n}_\lambda\vec{\sigma})\otimes|i\rangle\langle i|,
\end{equation}
where $\sigma_i$ are the Pauli matrices and the term $\lambda(\vec{n}_\lambda\vec{\sigma})$ describes a weak external field $\lambda\ll \omega_0$ in the direction of the unit vector $\vec{n}_\lambda$. 
We consider a system-bath Hamiltonian $H_{SB}$ of the following form,
\begin{equation}
H_{SB}=\sum_{i=1}^M g_{i,n}\sigma_-\otimes|i+1\rangle\langle i|\otimes a_{i,n}^\dag+g_{i,n}^*\sigma_+\otimes|i\rangle\langle i+1|\otimes a_{i,n},
\end{equation}
where $|M+1\rangle\equiv |1\rangle$.
Taking into account that the external field in the Hamiltonian $H_S$ is weak $\lambda\ll \omega_0$ the system-bath Hamiltonian in the interaction picture reads,
\begin{equation}
H_{SB}=\sum_{i=1}^M g_{i,n}\sigma_-\otimes|i+1\rangle\langle i|\otimes a_{i,n}^\dag e^{-i(\omega_0-\omega_{i,n})t}+\mathrm{h.c.}.
\end{equation}
The corresponding quantum master equation for the reduced density matrix takes the following form,
\begin{eqnarray}
\label{eq:2lvl}
\frac{d}{dt}\rho_s(t)&=&\sum_{i=1}^M\Big(-i\left[\lambda(\vec{n}_\lambda\vec{\sigma})\otimes|i\rangle\langle i|, \rho_s(t)\right]\\\nonumber
&+&\gamma_i \left(-\omega_0\right)\mathcal{L}\left(\sigma_-\otimes|i+1\rangle\langle i|\right)\rho_s(t)\\\nonumber
&+&\gamma_i \left(\omega_0\right) \mathcal{L}\left(\sigma_+\otimes|i\rangle\langle i+1|\right)\rho_s(t)\Big).
\end{eqnarray}
The superoperator defined by Eq. (\ref{eq:2lvl}) preserves the block-diagonal structure of the density matrix $\rho_s(t)=\sum_{i=1}^M\rho_i(t)\otimes|i\rangle\langle i|$. The continuous time OQWs has the same structure as Eq. (\ref{EQ:DMIJ1})
\begin{eqnarray}
\label{eq:2lvlCTOQW}
\frac{d}{dt}\rho_i(t)&=&-\imath\lambda\left[\vec{n}_\lambda\vec{\sigma}, \rho_i(t)\right]\\\nonumber
&+&\gamma_i\left(-\omega_0\right)\left(\sigma_-\rho_{i-1}(t)\sigma_+-\frac{1}{2}\{\sigma_+\sigma_-,\rho_i(t)\}_+\right)\\\nonumber
&+&\gamma_i\left(\omega_0\right)\left(\sigma_+\rho_{i+1}(t)\sigma_--\frac{1}{2}\{\sigma_-\sigma_+,\rho_i(t)\}_+\right).
\end{eqnarray}
Finally, it is easy to obtain the explicit form of the operators $B_i^j$ following the time dicretisation procedure introduced in Eq.(\ref{eq:2lvlCTOQW}),
\begin{eqnarray}
\label{eq:EX1}
B&=&\sqrt{\Delta\gamma \left(\langle n\rangle+1\right)}\sigma_-, C=\sqrt{\Delta\gamma\langle n\rangle}\sigma_+,\\\nonumber 
A&=&I_2-\frac{\Delta}{2}\left(\gamma \left(\langle n\rangle+1\right)\sigma_+\sigma_-+\gamma\langle n\rangle\sigma_-\sigma_+\right)-\imath \lambda\Delta\vec{n}_\lambda\vec{\sigma}.
\end{eqnarray}
For simplicity, in Eq. (\ref{eq:EX1}) we assume that all damping rates are the same, i.e. $\forall i, \gamma_i(-\omega_0)\equiv \gamma(-\omega_0)=\gamma\left(\langle n\rangle+1\right)$ and $\forall i,  \gamma_i(\omega_0)\equiv \gamma(\omega_0)=\gamma\langle n\rangle$, where $\gamma$ is the coefficient of spontaneous emission and $\langle n\rangle=\left(\exp{\left(\frac{\hbar\omega_0}{k_B T}\right)}-1\right)^{-1}$ is the mean number of thermal photons on the frequency $\omega_0$ for a bath at temperature $T$.
The iteration formula for OQWs with jump operators $A$, $B$ and $C$ (Eq.~\ref{eq:EX1}) is given by Eq.(\ref{eq:iteration}).
\begin{figure}
\includegraphics[width= \linewidth]{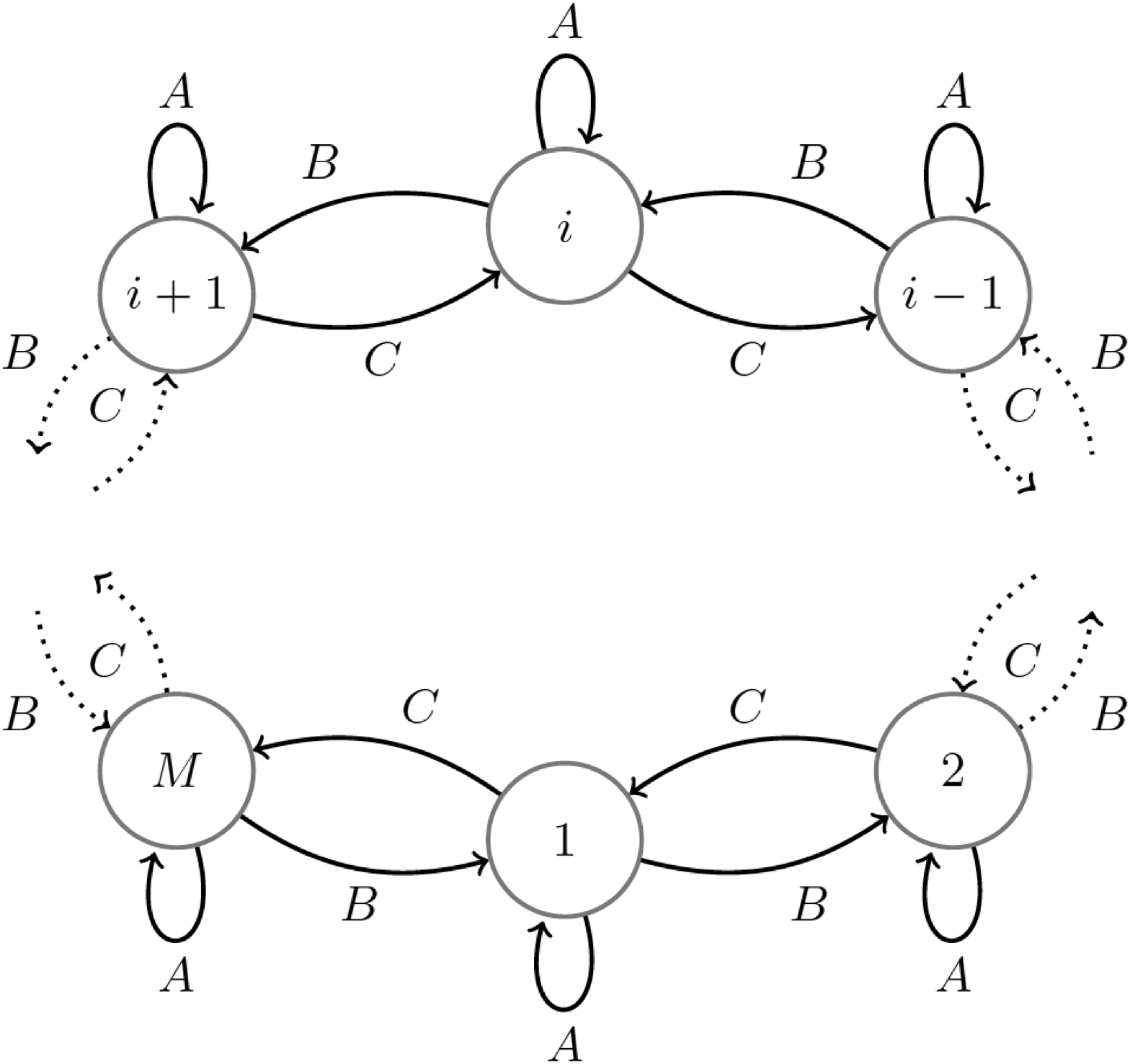}
\label{fig:DQRWZ1}
\caption{Open quantum walk on circle of nodes. A schematic representation of the OQW on circle: all transition to the right are induced by the operator $B$, while all transitions to the left are induced by the operator $C$ and all steps without transitions are induced by the operator $A$.}
\end{figure}

\begin{figure}
\includegraphics[width= \linewidth]{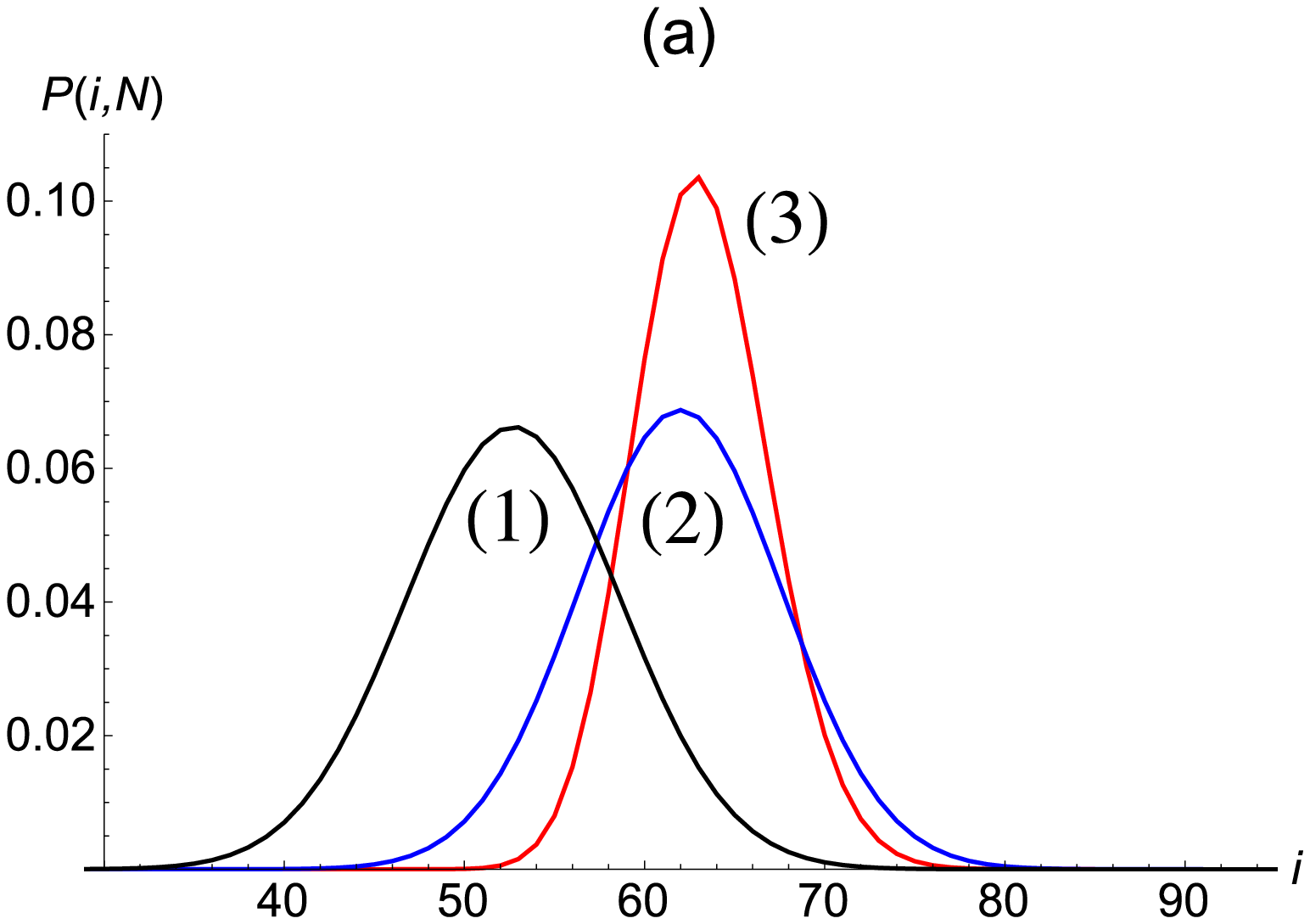}\\
\includegraphics[width= \linewidth]{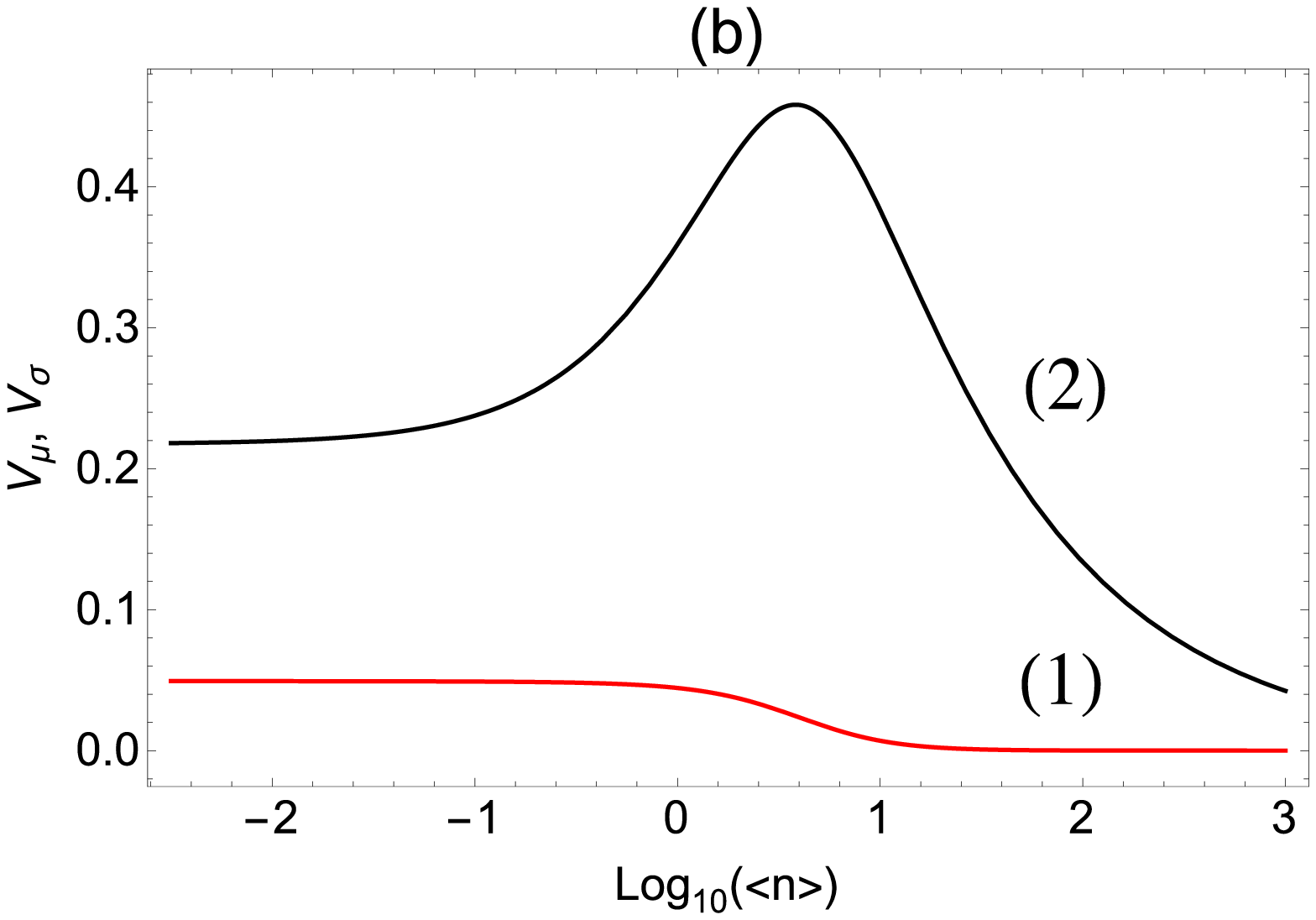}
\caption{(colour online) Open quantum walk on circle of nodes. Operators $A$, $B$ and $C$ are given by Eq. (\ref{eq:EX1}). Figure (a) shows the occupation probability of the ``walker" $P(i,N)=\mathrm{Tr}[\rho_i^{[N]}]$ after 5000 steps. The initial state of the ``walker" is given by $\rho^{[0]}=\frac{1}{2}I_2\otimes|51\rangle\langle 51|$ and curves (1)-(3) correspond to the different temperatures of the environment $\langle n\rangle=10, 1$ and $0.1$, respectively; the parameters are $\gamma=0.1$, $\lambda=0.3$ and $\Delta=0.05$. Figure (b) shows the dependence of the ``speed" of the gaussian $V_\mu$ (curve 1) and ``spread" of the gaussian $V_\sigma$ (curve 2) given by Eq. (\ref{vmu}) and Eq. (\ref{vs}) as function of the temperature of the environment; the parameters are $\gamma=0.1$ and $\lambda=0.3$.}
\end{figure}

\begin{figure}
\includegraphics[width= \linewidth]{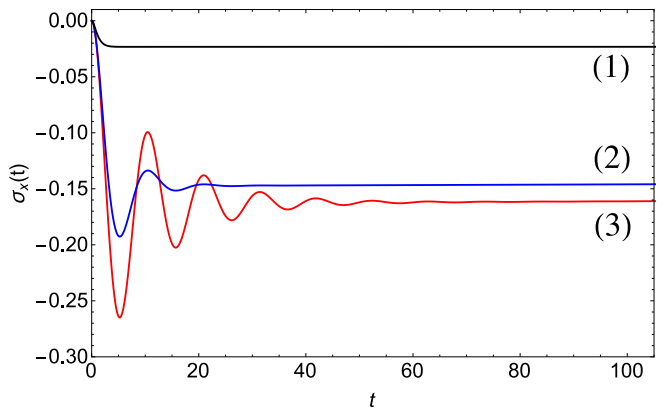}
\caption{(colour online) Open quantum walk on circle of nodes. Operators $A$, $B$ and $C$ are given by Eq. (\ref{eq:EX1}). The figure shows the dynamics of the total coherence of the ``walker" $\sigma_x(t)=\sum_{i=1}^{M}\mathrm{Tr}[\rho_i^{[t]}\sigma_x]$. The initial state of the ``walker" is given by $\rho^{[0]}=\frac{1}{2}I_2\otimes|51\rangle\langle 51|$ and the curves (1)-(3) correspond to the different temperatures of the environment $\langle n\rangle=10, 1$ and $0.1$, respectively; the parameters are $\gamma=0.1$, $\lambda=0.3$ and $\Delta=0.05$.}
\end{figure}

\begin{figure}
\includegraphics[width= \linewidth]{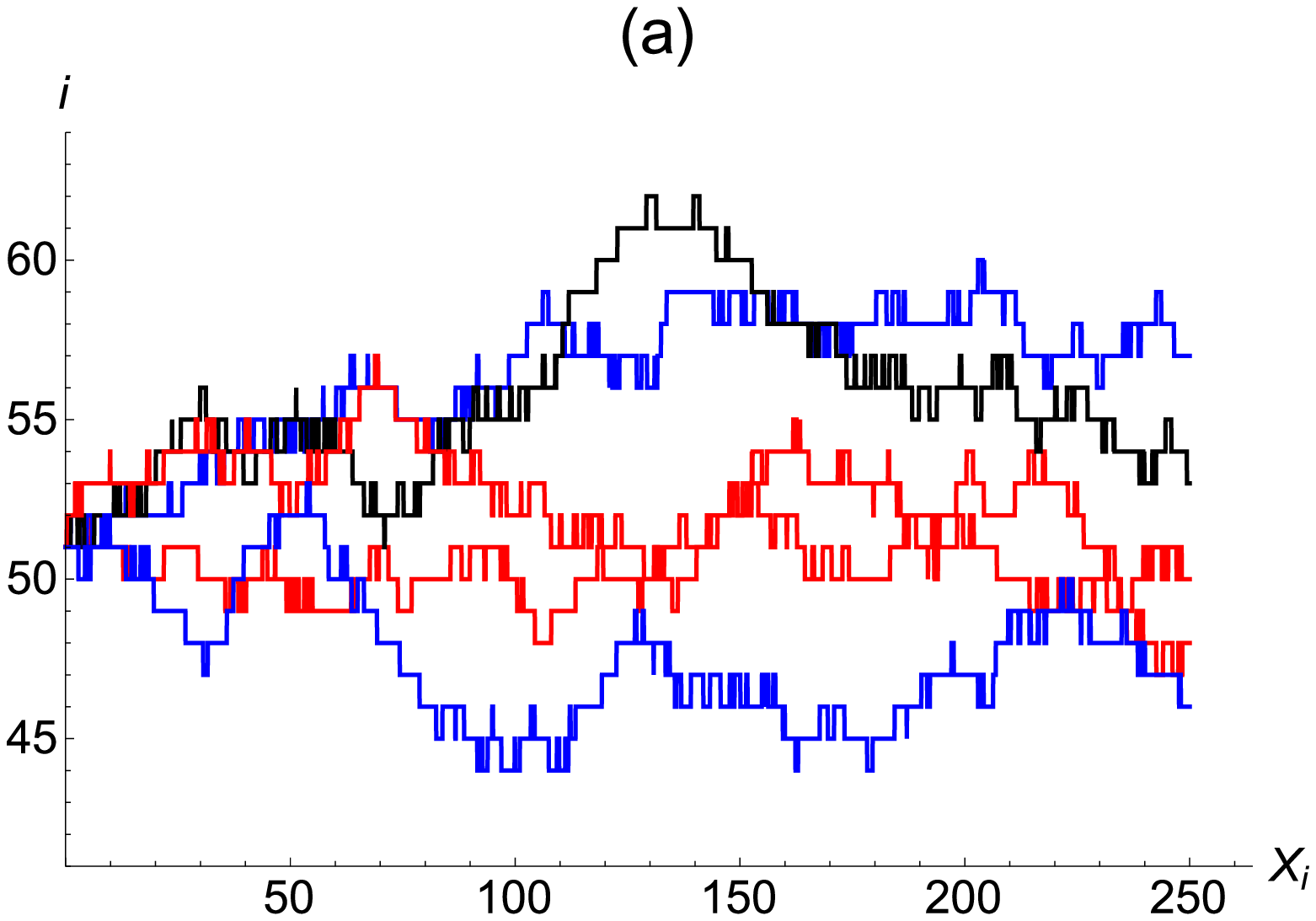}\\
\includegraphics[width= \linewidth]{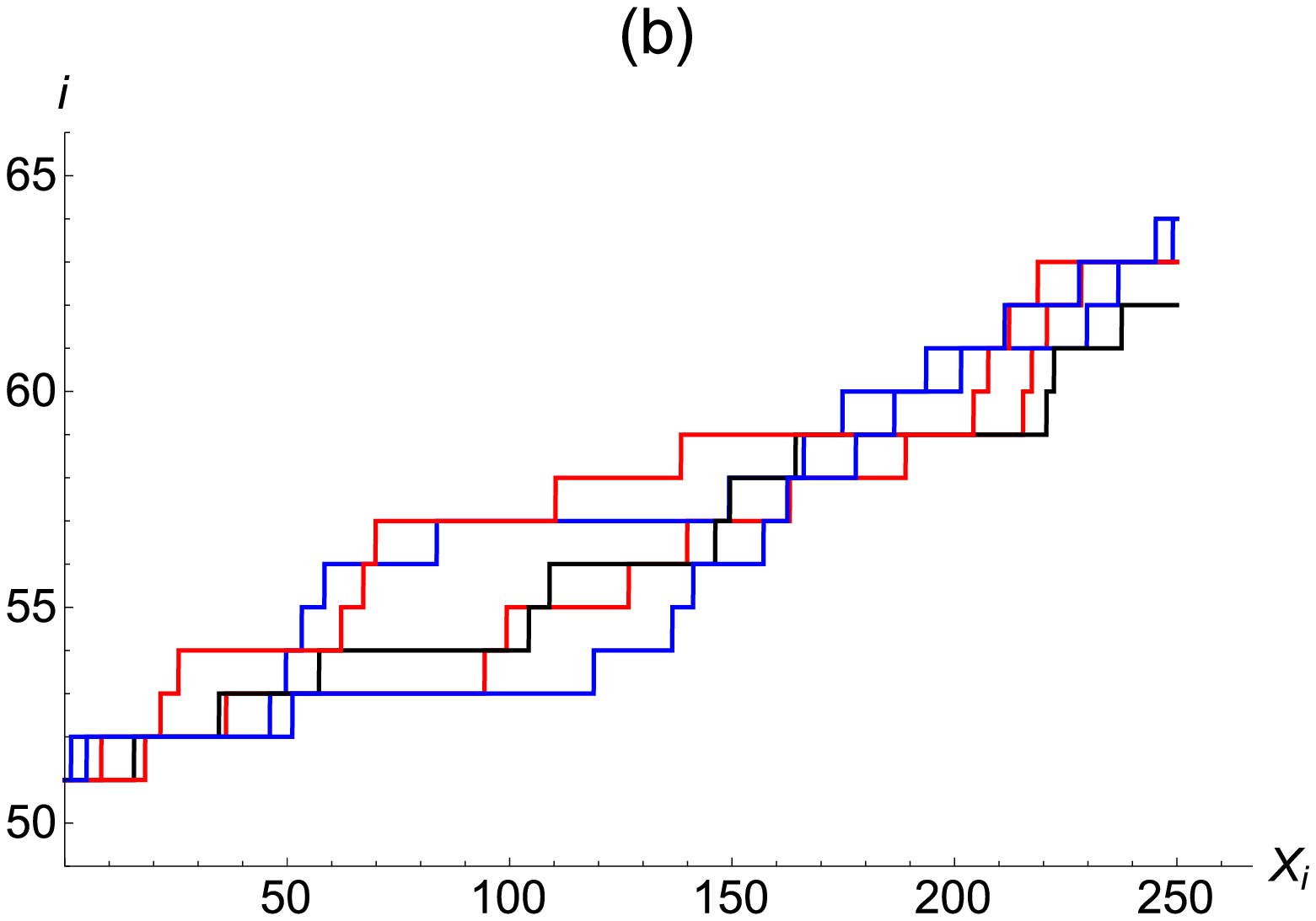}
\caption{(colour online) Open quantum walk on a circle of nodes. Operators $A$, $B$ and $C$ are given by Eq. (\ref{eq:EX1}). Figures (a) and (b) show examples of quantum trajectories of the ``walker" in the diffusive and the ballistic case, respectively. The diffusive behaviour of the quantum trajectories of the ``walker"  Fig. 5(a) corresponds to non-zero temperature of the environment $(\langle n\rangle=5)$, while the ballistic behaviour of the quantum trajectories of the ``walker"  Fig. 5(b) corresponds to an environment at zero temperature $(\langle n\rangle=0)$. The initial state of the ``walker" is given by $\rho^{[0]}=\frac{1}{2}I_2\otimes|51\rangle\langle 51|$; the parameters are $\gamma=0.1$, $\lambda=0.3$ and $\Delta=0.05$.}
\end{figure}

Fig. 3 shows the dynamics of different observables for the OQW on the circle with $M=101$ nodes and the jump operators given by Eq. (\ref{eq:EX1}). The occupation probability of the ``walker" $P(i,N)=\mathrm{Tr}[\rho_i^{[N]}]$ after $N_{\mathrm{steps}}=5000$ steps for different temperatures of environment is shown in Fig. 3(a). 
It is clear that decreasing the temperature of the bath (from Fig. 3(a1) to Fig.3(a3)) the gaussian distribution describing the occupation probability of the position of the ``walker" moves faster to the right and in the case of higher temperatures of the environment (Fig. 3(a1)) the average position of the ``walker" essentially remains near the initial node $51$. However, the width of the gaussian distributions corresponding to the temperatures $\langle n\rangle=1$ and $\langle n\rangle=10$ seems to be the same. To explain this one needs to analyse the average and variance of the position of the ``walker". Using Eq. (\ref{eq:2lvlCTOQW}) one can derive a system of equations for the mean $\mu(t) =\sum_{i=1}^Mi\mathrm{Tr}[\rho_i(t)]$ and the variance $\sigma^2(t)=\sum_{i=1}^M\left(i^2\mathrm{Tr}[\rho_i(t)]-\mu(t)^2\right)$ of the ``walker" in position space (see the Appendix for details). Using analytical expressions for $\mu(t)$ and $\sigma^2(t)$ it is possible to calculate the asymptotic velocity of this quantities,
\begin{equation}
\label{vmu}
V_\mu=\lim_{t\rightarrow\infty}\frac{\mu(t)}{t}=\frac{4\gamma\lambda^2}{\Omega^2},
\end{equation} 
where $\Omega=\sqrt{8\lambda^2+\gamma^2(2\langle n\rangle+1)^2}$, and
\begin{eqnarray}
\label{vs}
V_\sigma^2&=&\lim_{t\rightarrow\infty}\frac{\sigma^2(t)}{t}=\frac{\gamma}{2}\left(2\langle n\rangle+1\right)\\\nonumber
&-&\frac{3}{2}\frac{\gamma^7\left(2\langle n\rangle+1\right)^5}{\Omega^6}+\frac{3\gamma^5\left(2\langle n\rangle+1\right)^3}{\Omega^4}\\\nonumber
&-&\frac{\gamma^3\left(2\langle n\rangle+1\right)\left(\langle n\rangle^2+\langle n\rangle+1\right)}{\Omega^2}.
\end{eqnarray}

Fig. 3b shows the dependence of the ``speed" ($V_\mu$, curve 1) and ``spread" ($V_\sigma$, curve 2) of the gaussian distributions as a function of the temperature of the environment on a logarithmic scale. This figure perfectly explains the dynamics of the gaussians shown on Fig. 3a. Fig. 3b shows that the ``speed" of the gaussian is a monotonically decreasing function of the temperature of the environment. However, the biggest change in the velocity is happening for the temperature corresponding to the average number of photons in the bath between one ($\log_{10}\langle n\rangle=0$) and ten ($\log_{10}\langle n\rangle=1$). Fig. 3a is demonstrating that dependence: a gaussian distribution curve Fig. 3(a1) ($\log_{10}\langle n\rangle=1$) is much slower than Figs. 3(a2) and 3(a3) corresponding to $\log_{10}\langle n\rangle=0$ and $\log_{10}\langle n\rangle=-1$, respectively. The ``speed" ($V_\mu$) corresponding to Figs. 3(a2) and 3(a3) is approximately the same. The dependence of the ``spread" of the gaussians as a function of the temperature of environment (Fig. 3(b2)) is non-monotonic.With increasing of the temperature of the environment the ``spread" is growing to a certain point and afterwards decreasing. This non-monotonic dependence explains the same width of the gaussians corresponding to different temperatures of the bath Fig. 3(a1) ($\log_{10}\langle n\rangle=1$) and Fig. 3(a2) ($\log_{10}\langle n\rangle=0$).

Fig. 4 shows the dynamics of the total coherence
\begin{equation}
\sigma_x(t)=\sum_{i=1}^{M}\mathrm{Tr}[\rho_i^{[t]}\sigma_x]
\end{equation}
 of the ``walker". The time in Fig. 4 is in number of steps of the walk multiplied by the time - discretisation step $\Delta$, so each unit of time correspond to $20$ steps of the walk. One can see that for all considered temperatures of the environment there is some non-zero level of coherence present in the system. The presence of the steady-state coherence in the OQWs demonstrates that even if the steady-state position of the walker is a classically distributed degree of freedom, the inner state of the walker remains quantum. Obviously, for lower temperatures (Fig. 4(3)) the amount of coherence is higher and it takes more steps to achieve a steady-state coherence. Figs. 5 show examples of quantum trajectories of the ``walker" for non-zero and zero temperature of the environment, respectively. The quantum trajectories of the ``walker" were obtained using the unravelling of the OQWs \cite{pla,JSP,jpc}. In the zero temperature case the jump operator $C$ vanishes $(C\equiv 0)$, which implies that there will be only two options for the walker, namely to stay on the same node or to move to the right. The temperature of the bath plays the role of a switch between diffusive and ballistic trajectories of the ``walker".

Similar behaviour of the OQWs was described by Bauer {\it et. al} \cite{bm1} using parametrised generic CPTP maps. Only the microscopic derivation presented here allows to identify the physical conditions necessary to observe the transition in the behaviour of quantum trajectories.

\subsection{Second example}

\begin{figure}
\includegraphics[width= \linewidth]{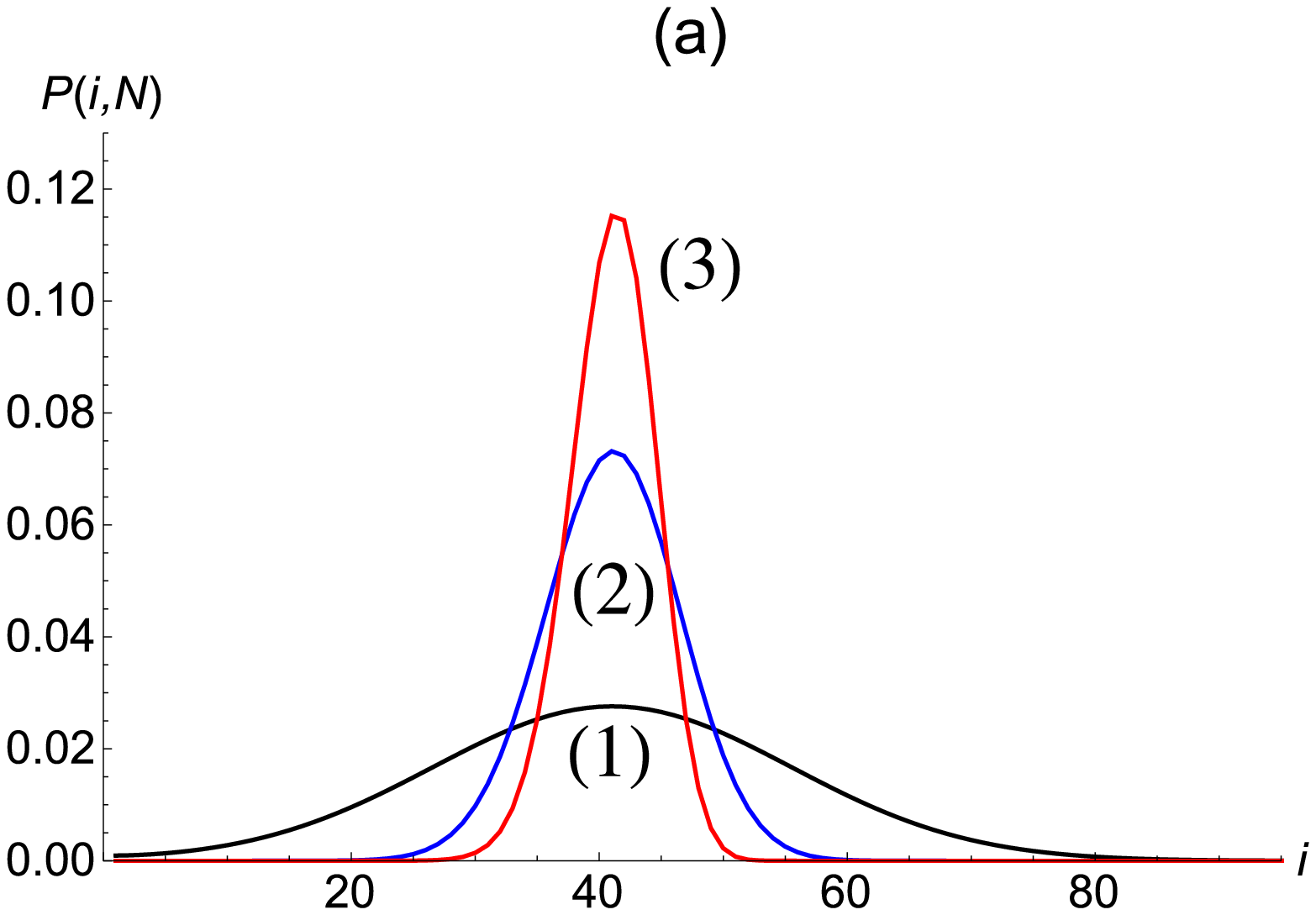}\\
\includegraphics[width= \linewidth]{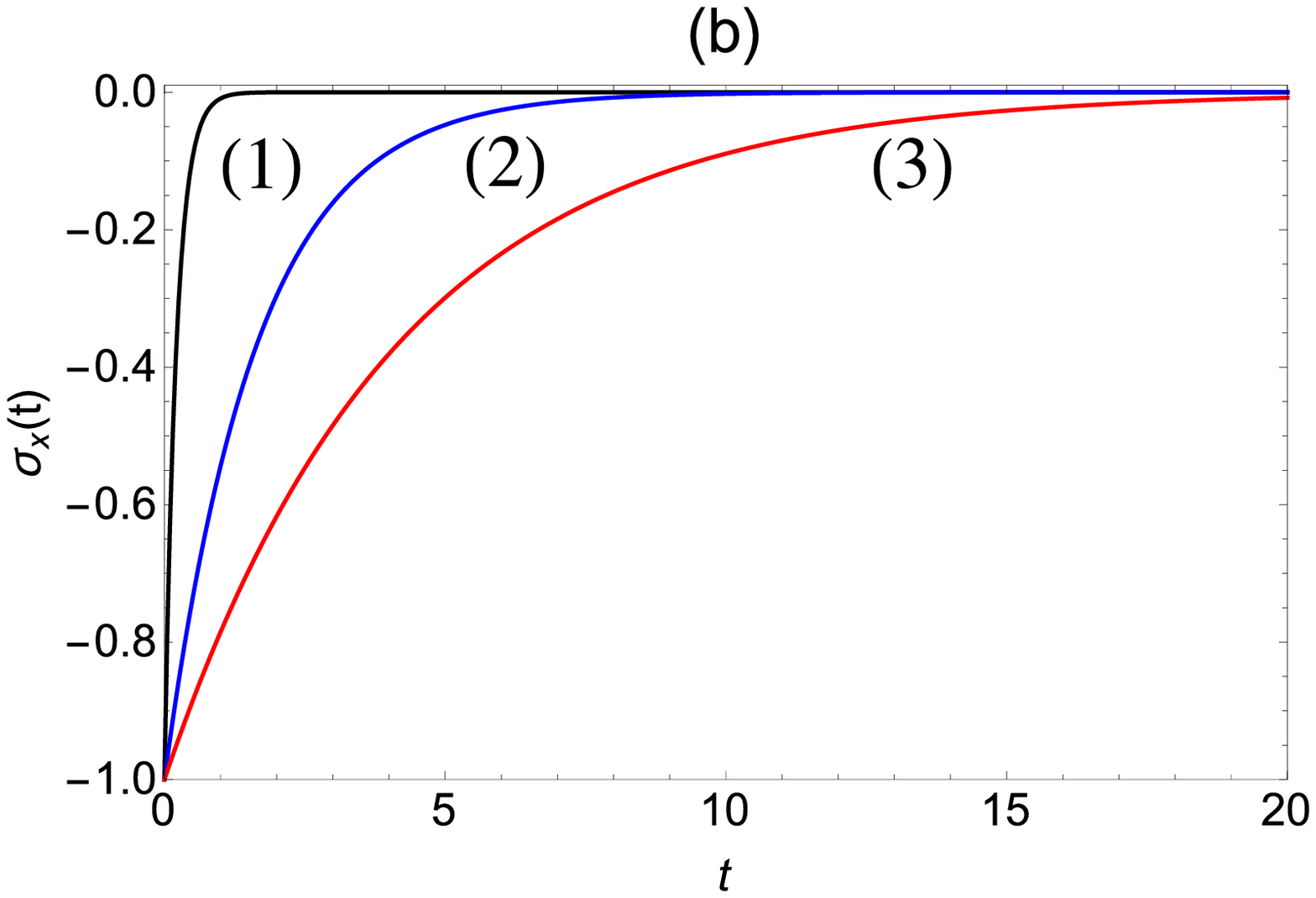}
\caption{(colour online) Open quantum walk on a finite chain of nodes. Figures (a) and (b) show the occupation probability of the ``walker" $P(i,N)=\mathrm{Tr}[\rho_i^{[N]}]$ after 5000 steps and the dynamics of the total coherence of the ``walker" $\sigma_x(t)=\sum_{i=1}^{M}\mathrm{Tr}[\rho_i^{[t]}\sigma_x]$. The initial state of the ``walker" is given by $\rho^{[0]}=|-\rangle\langle -|\otimes|51\rangle\langle 51|$ and the curves (1)-(3) correspond to the different temperatures of the environment $\langle n\rangle=10, 1$ and $0.1$, respectively; the parameters are $\gamma=0.1$, $\lambda=0.3$, $\Delta=0.05$, $\alpha=1$, $\beta=0$ and $|-\rangle=\frac{1}{\sqrt{2}}\left(|1\rangle-|0\rangle\right)$.}
\end{figure}

\begin{figure}
\includegraphics[width= \linewidth]{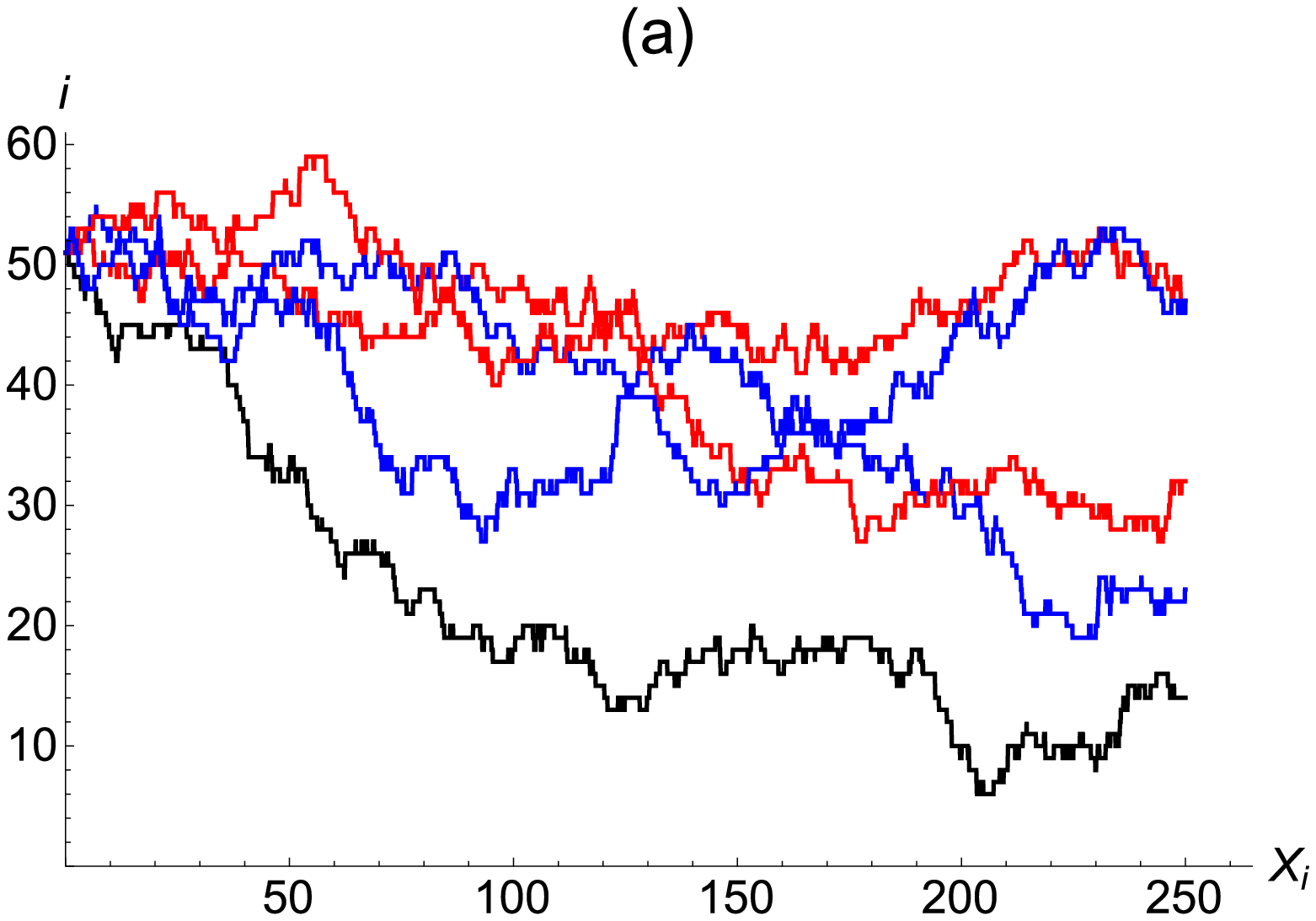}\\
\includegraphics[width= \linewidth]{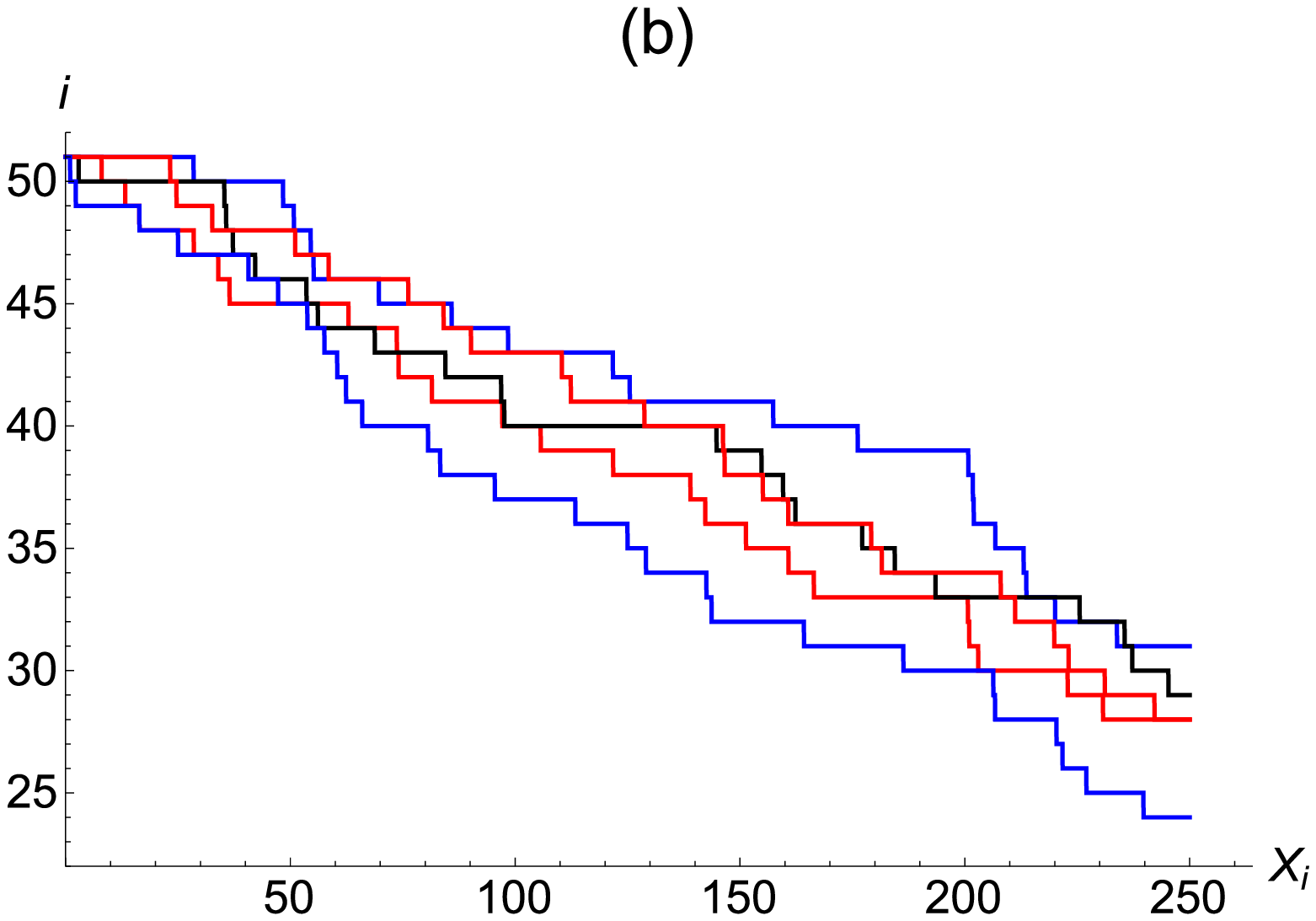}
\caption{(colour online) Open quantum walk on a finite chain of nodes. 
Figures (a) and (b) show examples of quantum trajectories of the ``walker" in the diffusive and the ballistic case, respectively. The diffusive behaviour of the quantum trajectories of the ``walker"  Fig. 7(a) corresponds to non-zero temperature of the environment $(\langle n\rangle=5)$, while the ballistic behaviour of the quantum trajectories of the ``walker"  Fig. 7(b) corresponds to an environment at zero temperature $(\langle n\rangle=0)$. The initial state of the ``walker" is given by $\rho^{[0]}=|-\rangle\langle -|\otimes|51\rangle\langle 51|$; the parameters are $\gamma=0.1$, $\lambda=0.3$, $\Delta=0.05$, $\alpha=1$, $\beta=0$ and $|-\rangle=\frac{1}{\sqrt{2}}\left(|1\rangle-|0\rangle\right)$.}
\end{figure}

The second example of the microscopic derivation is an OQW on a finite chain of nodes. As in the previous example the quantum walker will be a two level system. The Hamiltonian of the 
walker on the finite chain of nodes is given by,
\begin{equation}
H_S=\sum_{i=1}^{M}\frac{\epsilon_i}{2}\sigma_z\otimes| i \rangle\langle i|
\end{equation}
where the constants $\epsilon_k$ read,
\begin{equation}
\epsilon_k=\epsilon_0+k\Delta_0,
\end{equation}
where $\epsilon_0$ and $\Delta_0$ are some positive constants.
In this example we will assume a pure dephasing system-bath interaction,
\begin{equation}
H_{SB}=\sum_{i=1}^{M-1}\sum_n \left(\alpha \sigma_z+\beta I\right)X_{i,i+1}\otimes\left(g_{i,n}a_{i,n}+\mathrm{h.c.}\right),
\end{equation}
where $X_{i,i+1}=| i+1 \rangle\langle i|+| i \rangle\langle i+1|$ and $\alpha,\beta\in \mathbb{R}$. The free parameters $\alpha$ and $\beta$ allow us to address different types of dephasing coupling. For $\alpha=1, \beta=0$ we obtain $\sigma_z$ coupling and for $\alpha=\pm\beta=1/2$ we couple excited and ground levels of the two level system, respectively. In the interaction picture the Hamiltonian of the system bath interaction $H_{SB}(t)$ reads,
\begin{eqnarray}
H_{SB}(t)&=&\sum_{i=1}^{M-1}\sum_n \left(\alpha \sigma_z+\beta I\right)\otimes\big(| i+1 \rangle\langle i|e^{i\Delta_0 t}\\\nonumber
&+&| i \rangle\langle i+1|e^{-i\Delta_0 t}\big)\otimes\left(g_{i,n}a_{i,n}e^{-i\omega_{i,n} t}+\mathrm{h.c.}\right).
\end{eqnarray}
After application of the rotating wave approximation (RWA) we obtain the Hamiltonian of the system-bath interaction
\begin{eqnarray}
H_{SB}^{\mathrm{RWA}}(t)&=&\sum_{i=1}^{M-1}\sum_n \big(\alpha \sigma_z\\\nonumber&+&\beta I\big)\otimes\left(g_{i,n}| i \rangle\langle i+1|a_{i,n}e^{i(\omega_{i,n}-\Delta_0) t}+\mathrm{h.c.}\right).
\end{eqnarray}
It is straightforward to obtain the corresponding master equation in the Born-Markov approximation:
\begin{eqnarray}
\nonumber
\frac{d}{dt}\rho&=&\sum_{i=1}^{M-1}\gamma(\langle n\rangle+1)\mathcal{L}\Big(\left(\alpha \sigma_z+\beta I\right)\otimes| i \rangle\langle i+1|\Big)\rho\\
&+&\sum_{i=1}^{M-1}\gamma \langle n\rangle\mathcal{L}\Big(\left(\alpha \sigma_z+\beta I\right)\otimes| i+1 \rangle\langle i|\Big)\rho,
\end{eqnarray}
where $\gamma$ is the coefficient of spontaneous emission and $\langle n\rangle=\left(\exp{\left(\frac{\hbar\Delta_0}{k_B T}\right)}-1\right)^{-1}$ is the number of thermal photons on the frequency $\Delta_0$ of the bath at temperature  $T$.
As in the generic case Eq. (\ref{EQ:DMIJ}) and in the example Eq. (\ref{eq:2lvl}) the above master equation preserves the  block diagonal structure of the density matrix. By the direct substitution of $\rho_s(t)=\sum_{i=1}^M\rho_i(t)\otimes|i\rangle\langle i|$ one obtains the following generalised master equations:
\begin{equation}
\label{ME:SecEx}
\frac{d}{dt}\rho_1(t)=\gamma(\langle n\rangle+1) S\rho_2(t) S-\frac{\gamma \langle n\rangle}{2}\{S^2,\rho_1(t)\}_+,
\end{equation}
\begin{eqnarray}
\nonumber
\frac{d}{dt}\rho_i(t)&=&\gamma(\langle n\rangle+1) \left(S\rho_{i-1}(t) S-\frac{1}{2}\{S^2,\rho_i(t)\}_+\right)\\\nonumber
&+&\gamma\langle n\rangle \left(S\rho_{i+1}(t) S-\frac{1}{2}\{S^2,\rho_i(t)\}_+\right),\\\nonumber
& & (i=2,\dots,M-1),
\end{eqnarray}
\begin{equation}\nonumber
\frac{d}{dt}\rho_M(t)=\gamma\langle n\rangle S\rho_{M-1}(t) S-\frac{\gamma (\langle n\rangle+1)}{2}\{S^2,\rho_M(t)\}_+,
\end{equation}
where the operator $S$ is defined as, $S=S^\dag=\alpha \sigma_z+\beta I$.
Performing the discretisation procedure in Eq. (\ref{ME:SecEx}) we obtain the following jump operators:
\begin{equation}
\label{eq:EX2}
B_{i+1}^i=\sqrt{\gamma(\langle n\rangle+1)\Delta}S,
\end{equation}
\begin{equation}\nonumber
B_{i}^{i+1}=\sqrt{\gamma\langle n\rangle\Delta}S\quad (i=1,\dots,M-1),
\end{equation}
\begin{equation}\nonumber
B_{j}^{j}=I-\frac{\gamma(2\langle n\rangle+1)\Delta}{2}S^2\quad (j=2,\dots,M-1),
\end{equation}
\begin{equation}\nonumber
B_{1}^{1}=I-\frac{\gamma(\langle n\rangle+1)\Delta}{2}S^2,\quad B_{M}^{M}=I-\frac{\gamma\langle n\rangle\Delta}{2}S^2.
\end{equation}
The interation formula takes the following form,
\begin{eqnarray}
\rho_1^{[n+1]}&=&B_1^1\rho_{1}^{[n]}B_1^{1\dag}+B_2^1\rho_{2}^{[n]}B_2^{1\dag},\\\nonumber
\rho_M^{[n+1]}&=&B_M^M\rho_{M}^{[n]}B_M^{M\dag}+B_{M-1}^M\rho_{M-1}^{[n]}B_{M-1}^{M\dag},\\\nonumber
\rho_i^{[n+1]}&=&B_i^i\rho_{i}^{[n]}B_i^{i\dag}+B_{i-1}^i\rho_{i-1}^{[n]}B_{i-1}^{i\dag}+B_{i+1}^i\rho_{i+1}^{[n]}B_{i+1}^{i\dag},\\\nonumber
& &\left(i=2,\ldots,M-1\right).
\end{eqnarray}
Fig. 6 shows the dynamics of different observables of the OQW on the line with $M=101$ nodes and the jump operators given by Eq. (\ref{eq:EX2}). 
The occupation probability of the ``walker" $P(i,N)=\mathrm{Tr}[\rho_i^{[N]}]$ after $N_{\mathrm{steps}}=5000$ steps for different temperatures of environment is shown in Fig. 6(a). As in the previous example  decreasing the temperature of the bath (from Fig. 6(a1) to Fig.6(a3)) the dispersion of the distribution grows. However, in contrast to the previous example the average speed of the ``walker" is independent of the temperature of the bath. In this special example the jump operators ($A$, $B$ and $C$) are diagonal, this implies that each operator $\rho_i^{[n]}$ ($\rho_{i(k)}^{[n]}=\langle k|\rho_i^{[n]}|k\rangle$) evolves independently from other elements of the operator $\rho_i^{[n]}$. The iteration formula for the diagonal elements $\rho_{i(k)}^{[n]}$ reads,
\begin{eqnarray}
\label{eq:EX21}
\rho_{i(k)}^{[n+1]}&=&\left(1-\gamma\left(2\langle n\rangle+1\right)\Delta\right)\rho_{i(k)}^{[n]}\\\nonumber
&+&\gamma\left(\langle n\rangle+1\right)\Delta\rho_{i+1(k)}^{[n]}+\gamma\langle n\rangle\Delta\rho_{i-1(k)}^{[n]}.
\end{eqnarray}
This means that on average the ``walker" is always moving to the left with drift velocity $\gamma\Delta$ (Fig. 6(a)). Fig. 6(b) shows the dynamics of the total coherence of the ``walker" $\sigma_x(t)=\sum_{i=1}^{M}\mathrm{Tr}[\rho_i^{[t]}\sigma_x]$. Due to the pure decoherent interaction with the heat bath one can see that for all temperatures the coherence is vanishing. As it is expected for higher temperatures (Fig. 6(b1)) the coherence vanishes faster than for lower ones (Fig. 6(b3)). Fig. 7 show samples of the quantum trajectories of the "walker" for non-zero and zero temperature of the environment, respectively. For the zero temperature of the bath the jump operator $B_i^{i+1}$ vanishes $(B_i^{i+1}\equiv 0)$, which implies that the ``walker" can stay on the same node or move to the left. The continuous time limit $(\Delta\rightarrow0)$ of the iteration equation (\ref{eq:EX21}) is a differential equation for the Poisson process, i.e. $d\rho_{i(k)}/dt=\gamma\left(\rho_{i+1(k)}-\rho_{i(k)}\right)$.
As in the previous example, the temperature of the bath plays the role of a switch between diffusive and ballistic trajectories of the ``walker".

\vspace{1cm}

\section{Conclusion}

In this paper we presented the generic case of the microscopic derivation of Open Quantum Walks. We start by identifying the Hamiltonian of the ``walker" and the nodes as system Hamiltonian, the Hamiltonian of the reservoirs and the Hamiltonian of system-bath interaction. The Hamiltonian of the system-bath interaction was chosen such that only dissipative interaction drove transitions between the nodes of the OQW. We applied the Born-Markov approximation to obtain the reduced master equation of the density matrix of the system (``walker" and nodes of the walk). The resulting master equation has the generalised master equation form and defined a continuous time Open Quantum Walk. The generalised master equation conserves the diagonal in position form of the reduced density matrix. The time discretisation of the generalised master equation leads to discrete time OQW formalism.

The formalism was demonstrated for the examples of the OQW on a circle of nodes and an finite chain of nodes. The presented microscopic derivation allows to connect different dynamical behaviour of the OQWs with thermodynamical parameters of the total system. For both examples a transition between the diffusive and ballistic quantum trajectories was observed. The temperature of the bath was identified as switching parameter between these different types of the quantum trajectories. As it was shown in the example of the OQWs on a circle of nodes, there is a persistent coherence in the OQWs even after system reaches steady Gaussian distribution. 

It was shown that OQWs can efficiently performs dissipative quantum computing \cite{qip} and efficiently transport excitations \cite{pla}. However, only with help of a microscopic derivation it is possible to identify the Hamiltonians which are necessary to implement these walks in a realistic physical systems. This will be the subject of future research in this field.

\begin{acknowledgments}
This work is based upon research supported by the South African
Research Chair Initiative of the Department of Science and
Technology and National Research Foundation.
\end{acknowledgments}

\appendix
\section{Derivation of the coefficients $V_\mu$ and $V_\sigma^2$}

To derive an equation for the coefficient $V_\mu$ one needs to obtain the equation for $\mu(t) =\sum_{i=1}^Ni\mathrm{Tr}[\rho_i(t)]$.
The generalised master equation Eq. (\ref{eq:2lvlCTOQW}) can be rewritten as,
\begin{eqnarray}
\dot{Y}_i&=&-\frac{\gamma(2\langle n\rangle+1)}{2}Y_i,\\\nonumber
\dot{X}_i&=&2\lambda Z_i-\frac{\gamma(2\langle n\rangle+1)}{2}X_i,\\\nonumber
\dot{Z}_i&=&-2\lambda X_i -\frac{\gamma}{2}P_i-\frac{\gamma(2\langle n\rangle+1)}{2}Z_i\\\nonumber
&+&\frac{\gamma \langle n\rangle}{2}\left(P_{i+1}-Z_{i+1}\right)-\frac{\gamma(\langle n\rangle+1)}{2}\left(P_{i-1}+Z_{i-1}\right),\\\nonumber
\dot{P}_i&=&-\frac{\gamma(2\langle n\rangle+1)}{2}P_i-\frac{\gamma}{2}Z_i\\\nonumber
&+&\frac{\gamma(\langle n\rangle+1)}{2}\left(P_{i-1}+Z_{i-1}\right)+\frac{\gamma \langle n\rangle}{2}\left(P_{i+1}-Z_{i+1}\right),
\end{eqnarray}
where the index $i$ runs from $1$ to $M$ with periodic boundary conditions ($M+1\equiv 1$). The functions $Y_i\ldots P_i$ are defined as, $P_i=\mathrm{Tr}[\rho_i(t)]$ and $A_i=\mathrm{Tr}[\sigma_A\rho_i(t)]$ ($\sigma_A$ is corresponding Pauli matrix). Using this system of differential equation for  $Y_i\ldots P_i$ one can find a corresponding system of differential equation for the following collective functions: $A_s=\sum_{i=1}^M A_i$, $\langle A\rangle= \sum_{i=1}^M i A_i$ and $\langle\langle A\rangle\rangle= \sum_{i=1}^M i^2 A_i$, where  $A_i \in (P_i, X_i, Y_i, Z_i)$. By definition the coefficients $V_\mu$ and $V_\sigma^2$ are asymptotic linear parts of the following functions $\mu(t) =\langle P\rangle$ and $\sigma^2(t)=\langle \langle P\rangle\rangle-\langle P\rangle^2$, respectively. 

Using the definition of the collective variable $A_s=\sum_{i=1}^M A_i$ and periodic boundary conditions it is easy to obtain a system of differential equations for $P_s$, $X_s$ and $Z_s$,
\begin{eqnarray}
\label{eq:AS}
\frac{d}{dt}P_s&=&0,\\\nonumber
\frac{d}{dt}\left(\begin{matrix} Z_s \\ X_s\end{matrix}\right)&=&G_2\left(\begin{matrix} Z_s \\ X_s\end{matrix}\right)-\left(\begin{matrix} \gamma P_s \\ 0\end{matrix}\right),
\end{eqnarray} 
where $G_2=\left(\begin{matrix} -\gamma(2\langle n\rangle+1) & -2\lambda \\ 2\lambda & -\gamma\frac{2\langle n\rangle+1}{2}\end{matrix}\right)$. The first equation of this system has a very simple physical meaning, knowing that $P_S=\sum_{i=1}^M P_i$, where $P_i$ is the probability to find the walker on the node $i$ and $\sum_{i=1}^M P_i$ is just the trace of the total reduced density matrix of the walker on the ring of nodes. This implies that $P_s$ is the total probability to find the walker on one of the nodes, so $P_s=P_s(0)=1$. The equation for $X_s$ and $Z_s$ can be integrated as,
\begin{equation}
\left(\begin{matrix} Z_s \\ X_s\end{matrix}\right)=e^{tG_2}\left(\begin{matrix} Z_s(0) \\ X_s(0)\end{matrix}\right)-\gamma\int_0^t d\tau e^{(t-\tau)G_2}\left(\begin{matrix} 1 \\ 0\end{matrix}\right).
\end{equation}

Now, we can write down the equation for $\langle A\rangle= \sum_{i=1}^M i A_i$,
\begin{eqnarray}
\label{eq:MA}
\frac{d}{dt}\langle P\rangle&=&\gamma\frac{2\langle n\rangle+1}{2}Z_s+\frac{\gamma}{2}P_s,\\\nonumber
\frac{d}{dt}\left(\begin{matrix} \langle Z\rangle \\ \langle X\rangle\end{matrix}\right)&=&G_2 \left(\begin{matrix} \langle Z\rangle \\ \langle X\rangle\end{matrix}\right)\\\nonumber&-&\left(\gamma \langle P\rangle+\frac{\gamma}{2}Z_s+\frac{\gamma(2\langle n\rangle+1)}{2}P_s\right)\left(\begin{matrix} 1 \\ 0\end{matrix}\right).
\end{eqnarray}
The formal solution for the function $\langle P\rangle$ has the form,
\begin{equation}
 \langle P\rangle(t)= \langle P\rangle(0)+\frac{\gamma t}{2} + \gamma\frac{2\langle n\rangle+1}{2}\int_0^t d\tau Z_s(\tau).
\end{equation}
Using the spectral decomposition theorem for the matrix $G_2$ it is easy to obtain,
\begin{eqnarray}
G_2&=&\lambda_+\Pi_++\lambda_-\Pi_-,\\\nonumber
\lambda_\pm&=&-\frac{3}{4}\gamma\left(2\langle n\rangle+1\right)\pm\frac{\omega}{4},\\\nonumber
\Pi_\pm&=&\pm\frac{2}{\omega}\left(G_2-\lambda_\mp 1_2\right),
\end{eqnarray}
where $\lambda_\pm$ and $\Pi_\pm$ are eigenvalues and corresponding orthogonal projectors on the eigenspaces of the matrix $G_2$, $1_2$ - denotes $2\times 2$ identity matrix and the constant $\omega$ is given by $\omega=\sqrt{\gamma^2\left(2\langle n\rangle+1\right)^2-64\lambda^2}$.
Using the decomposition of the matrix $G_2$ the expression for the collective variable $\langle P\rangle$ reads,
\begin{eqnarray}
\label{eqprev01}
\langle P\rangle(t)&=&\langle P\rangle(0)+\frac{\gamma t}{2}\\\nonumber
&+&\gamma\frac{2\langle n\rangle+1}{2}\sum_{k=+,-}\Big\{\frac{e^{\lambda_k t}-1}{\lambda_k} \left(\begin{matrix} 1 & 0\end{matrix}\right)\Pi_k\left(\begin{matrix} Z_s(0) \\ X_s(0)\end{matrix}\right)\\\nonumber
&-&\frac{\gamma}{\lambda_k^2}\left(e^{\lambda_k t}-1 -\lambda_k t\right)\left(\begin{matrix} 1 & 0\end{matrix}\right)\Pi_k\left(\begin{matrix} 1 \\ 0\end{matrix}\right)\Big\}.
\end{eqnarray}
With the help of Eq. (\ref{eqprev01}) the coefficient $V_\mu$ is given by,
\begin{eqnarray}
\label{avmu}
V_\mu&=&\lim_{t\rightarrow\infty}\frac{\langle P\rangle(t)}{t}=\frac{\gamma}{2}\\\nonumber
&+&\gamma^2\frac{2\langle n\rangle+1}{2}\sum_{k=+,-}\frac{1}{\lambda_k}\left(\begin{matrix} 1 & 0\end{matrix}\right)\Pi_k\left(\begin{matrix} 1 \\ 0\end{matrix}\right)\\\nonumber
&=&\frac{4\gamma\lambda^2}{8\lambda^2+\gamma^2(2\langle n\rangle+1)^2}.
\end{eqnarray}

The differential equation for the collective variable $\langle\langle P\rangle\rangle$ has the form,
\begin{equation}
\frac{d}{dt}\langle\langle P\rangle\rangle=\gamma\langle P\rangle+\gamma\left(2\langle n\rangle+1\right)\langle Z\rangle+\frac{\gamma}{2}Z_s+\frac{\gamma\left(2\langle n\rangle+1\right)}{2}.
\end{equation}
The solution of the function $\langle Z\rangle$ is obtained from the system (\ref{eq:MA}) as,
\begin{eqnarray}
\left(\begin{matrix} \langle Z\rangle \\ \langle X\rangle\end{matrix}\right)&=&e^{tG_2}\left(\begin{matrix} \langle Z\rangle(0) \\ \langle X\rangle(0)\end{matrix}\right)
-\gamma\int_0^t d\tau \Big\{ \langle P\rangle(\tau)\\\nonumber
&+&\frac{1}{2}Z_s(\tau)+\frac{2\langle n\rangle+1}{2}\Big\}e^{(t-\tau)G_2}\left(\begin{matrix} 1 \\ 0\end{matrix}\right).
\end{eqnarray}
Using the explicit solutions for the functions $ \langle Z\rangle$, $ \langle P\rangle$ and $Z_s$ it is straightforward to obtain the solution for the function $\langle\langle P\rangle\rangle$. The coefficient $V_\sigma^2$ can be obtained as,
\begin{eqnarray}
\label{avs}
V_\sigma^2&=&\lim_{t\rightarrow\infty}\frac{\langle\langle P\rangle\rangle-\langle P\rangle^2}{t}=\frac{\gamma}{2}\left(2\langle n\rangle+1\right)\\\nonumber
&-&\frac{3}{2}\frac{\gamma^7\left(2\langle n\rangle+1\right)^5}{\Omega^6}+\frac{3\gamma^5\left(2\langle n\rangle+1\right)^3}{\Omega^4}\\\nonumber
&-&\frac{\gamma^3\left(2\langle n\rangle+1\right)\left(\langle n\rangle^2+\langle n\rangle+1\right)}{\Omega^2},
\end{eqnarray}

where $\Omega=\sqrt{8\lambda^2+\gamma^2(2\langle n\rangle+1)^2}$.

The expressions (\ref{avmu}) and (\ref{avs}) conclude the derivation of Eqs. (\ref{vmu},\ref{vs}), respectively. 


\end{document}